\documentclass[twocolumn,twoside,9pt]{extarticle}
\usepackage{amssymb}
\usepackage{enumerate}
\usepackage{a4wide}
\usepackage[english]{babel}
\usepackage{amsmath}
\usepackage{graphicx,algorithm,algorithmic,url,hyperref,epstopdf,authblk}
\usepackage{caption3} % load caption package kernel first
\DeclareCaptionOption{parskip}[]{} % disable "parskip" caption option
\usepackage[small]{caption}

\begin{document}
%\firstpage{1}

\title{SASeq: A Selective and Adaptive Shrinkage Approach to Identify and Quantify Condition-Specific Transcripts using RNA-Seq}
\author{Tin Chi Nguyen\,$^{1}$, Nan Deng\,$^{1}$ and Dongxiao Zhu\,$^{1,*}$}
\affil{$^{1}$Department of Computer Science, Wayne State University,
         5057 Woodward Avenue, Detroit, MI 48202, USA.\\
        $^{*}$To whom correspondence should be addressed.}
\date{}
\maketitle

\section*{Abstract}
Identification and quantification of condition-specific transcripts using RNA-Seq is vital in transcriptomics research. While initial efforts using mathematical or statistical modeling of read counts or per-base exonic signal have been successful, they may suffer from model overfitting since not all the reference transcripts in a database are expressed under a specific biological condition. Standard shrinkage approaches, such as Lasso, shrink all the transcript abundances to zero in a non-discriminative manner. Thus it does not necessarily yield the set of condition-specific transcripts. Informed shrinkage approaches, using the observed exonic coverage signal, are thus desirable. Motivated by ubiquitous uncovered exonic regions in RNA-Seq data, termed as ``naked exons", we propose a new computational approach that first filters out the reference transcripts not supported by splicing and paired-end reads, then followed by fitting a new mathematical model of per-base exonic coverage signal and the underlying transcripts structure. We introduce a tuning parameter to penalize the specific regions of the selected transcripts that were not supported by the naked exons. Our approach compares favorably with the selected competing methods in terms of both time complexity and accuracy using simulated and real-world data. Our method is implemented in SAMMate, a GUI software suite freely available from \url{sammate.sourceforge.net}.

\section{Introduction}
Identification and quantification of active transcripts using gene expression data is essential to a wide range of transcriptomics research. The problem is non-trivial since the observed exonic expression signal may be aggregated from a set of expressed transcripts encoded by the same gene via alternative splicing mechanisms. Moreover, excessive sequencing errors and biases exist in the data making the problem even more challenging~\cite{Roberts11,Wu2011,BNbiasCorrection12,WeiLi2012}. In earlier studies, several computational approaches have been developed to utilize high throughput gene expression profiling data collected from microarray experiments such as an Expectation-Maximization (EM) algorithm using Expressed Sequence Tags (ESTs)~\cite{Xing2006},
and a Nonnegative Matrix Decomposition (NMF) based algorithm~\cite{Anton2008} using exon and exon-exon junction microarrays. These approaches represent the pioneering efforts to tackle the isoform quantification problem. However, their performances were limited by data quantity (ESTs) and quality (microarrays).

With the widespread proliferation of next generation sequencing, unprecedented opportunities now exist for both genomics~\cite{Wheeler2008,Medvedev2009,Wade2009,Medvedev2010,Wu2012} and transcriptomics research~\cite{Mortazavi08,Wilhelm2008,Lister2008,Wang2009,Pepke2009,Tarazona2011,ManuelGarber2011}. Particularly, RNA-Seq technology has substantially improved both the data quantity and quality to better identify and quantify condition-specific transcripts. Existing approaches are either based on statistical modeling of read counts, such as Poisson~\cite{Jiang2009,Wang2010,Li2010,Trapnell2010,POME}, Gaussian~\cite{IsoInfer}, Multinomial~\cite{Nicolae2011,Deng2011}, and Dirichlet-Multinomial~\cite{Katz2010}, or are based on mathematical modeling of the relationships between per-base exonic coverage signal and transcript structures~\cite{Bohnert2010,DBLP:conf/bibm/NguyenDXDZ11}. Despite initial success, significant challenges remain in model overfitting. For example, there are around $130,000$ reference transcripts corresponding to some $20,000$ annotated human protein-coding genes in the Ensembl database (version 65). $59\%$ of these genes have $4$ or more annotated transcripts. However, only $1$ major transcript isoform and $1$ or $2$ minor transcript isoforms are typically expressed under a single biological condition, making the total number of condition-specific transcripts per gene less than $4$. Thus, model overfitting is inevitable for the majority of genes and appropriate model selection is urgently needed.

Existing shrinkage approaches, such as IsoLasso~\cite{IsoLasso} and SLIDE~\cite{LiJJ2011}, identify condition-specific transcripts by shrinking the abundance proportion parameters associated with each transcript isoform towards zero using the standard Lasso procedure~\cite{Tibshirani94}. They represent the first efforts to address model overfitting issues in identifying and quantifying condition-specific transcripts. Although their statistical models differ in details, they both divide the coding sequence into sub-exons. Read counts falling into these sub-exons are then used to estimate the transcript proportions. The approaches are limited since thousands of per-base signal are interpolated to only a dozen of data points, which are then used to estimate transcript expression. Furthermore, we argue that a straightforward application of a Lasso shrinkage approach for identifying condition-specific transcripts may be limited for a variety of reasons. First, choosing the tuning parameter for the Lasso shrinkage is not straightforward since the number of expressed transcripts is unknown and varies from gene to gene and from condition to condition. Second, the Lasso approach shrinks all the component coefficients, i.e. transcript abundance parameters, in a non-discriminative way and does not necessarily lead to the set of expressed transcripts~\cite{Tibshirani94,Leng2006,PengZhao2006,HuiZou2006}.

\begin{figure}[t]
\centering
\includegraphics[width=0.48\textwidth]{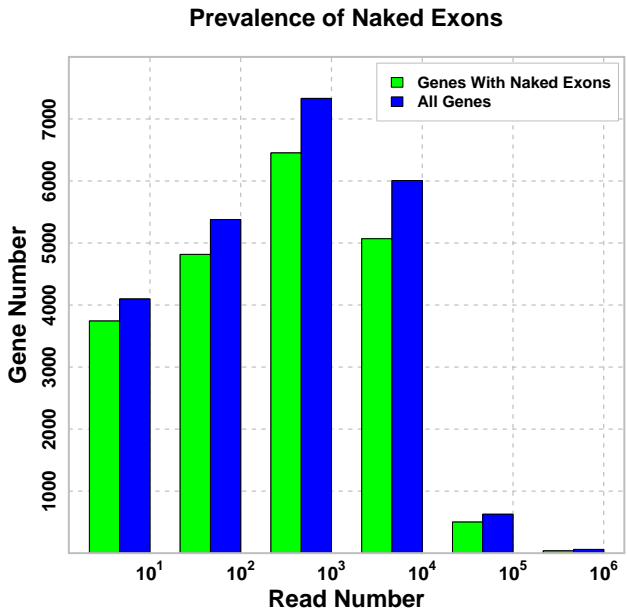}
\caption{Naked exons is prevalently observed in RNA-Seq read alignment results. The histogram's horizonal axis bins each gene based on the number of mapped reads. The histogram's vertical axis displays the number of genes falling into each bin. Blue corresponds to the total number of genes falling into each bin whereas green corresponds to a subset of genes with at least $10\%$ naked exons. The presented data set consists of $80$ million reads with length $75$. $20,617$ out of the total $23,511$ expressed genes ($88\%$) have naked exons.}
\label{fig:UncoveredRegions}
\end{figure}

Here we present a new computational approach that exploits both read counts and per-base exonic coverage signal, which carry alignment information from different aspects. Therefore, combining them is expected to yield a more accurate estimation of expression abundance. First, we use the reads that are compatible for each isoform to filter out potentially inactive transcript isoforms. Second, we develop a new Selective and Adaptive Shrinkage approach (SASeq) to address model overfitting issues using per-base exonic coverage signal. The rationale of our informed shrinkage approach is that transcripts with large naked exons in a highly expressed gene are likely to be inactive. Further we note the naked exons are prevalently observed in real-world RNA-Seq read alignment results, such as the one shown in Figure \ref{fig:UncoveredRegions}. Instead of shrinking all of the transcript abundance parameters, only selected regions in selected transcripts that are not supported by the naked exons are penalized. We employ a tuning parameter to penalize those regions. The tuning parameter represents the shrinkage level that is adjusted adaptively proportional to the exonic coverage via an optimization algorithm. We implement the new computational approach in our GUI software suite, SAMMate, which is freely available from \url{sammate.sourceforge.net}.

\section{Methods}
RNA-Seq reads can be aligned using many tools including TopHat~\cite{TopHat}, SplitSeek~\cite{SplitSeek}, MapSplice~\cite{WangKai2010}, ABMapper~\cite{ABMapper}, and SpliceMap~\cite{splicemap}. The alignment results are saved in SAM or BAM format that can be used as input for SASeq. For each gene, we extract the read-isoform compatibility information and also build a vector of per-base exonic signal from the alignment results.

\subsection{Pre-processing: Filtering and Signal Augmentation}
Each gene consists of a set of annotated exons whereas each transcript isoform consists of a subset of those exons. A single-end read is compatible with an isoform if it maps to one exon or if it contiguously spans across multiple exons of the isoform. A paired-end read is compatible with an isoform as long as both reads are compatible. As opposed to single-end reads, paired-end reads usually span a longer range and can thus help to identify longer splicing patterns. From the read-isoform compatibility information, existing read counts based approaches~\cite{Jiang2009,Wang2010,Li2010,Trapnell2010,POME,IsoInfer,Nicolae2011,Deng2011,Katz2010,IsoLasso,LiJJ2011} typically calculate the read counts falling into sub-exons and then infer the isoform proportions via iterative optimization algorithms. As the number of isoforms increases, the read counts falling into the sub-exons might be insufficient for transcriptome quantification purposes. Therefore, we use read-isoform compatibility information only to filter out inactive isoforms as opposed to calculating an exact numerical value for each isoform proportion.

\begin{figure*}
\centering
\includegraphics[width=0.96\textwidth]{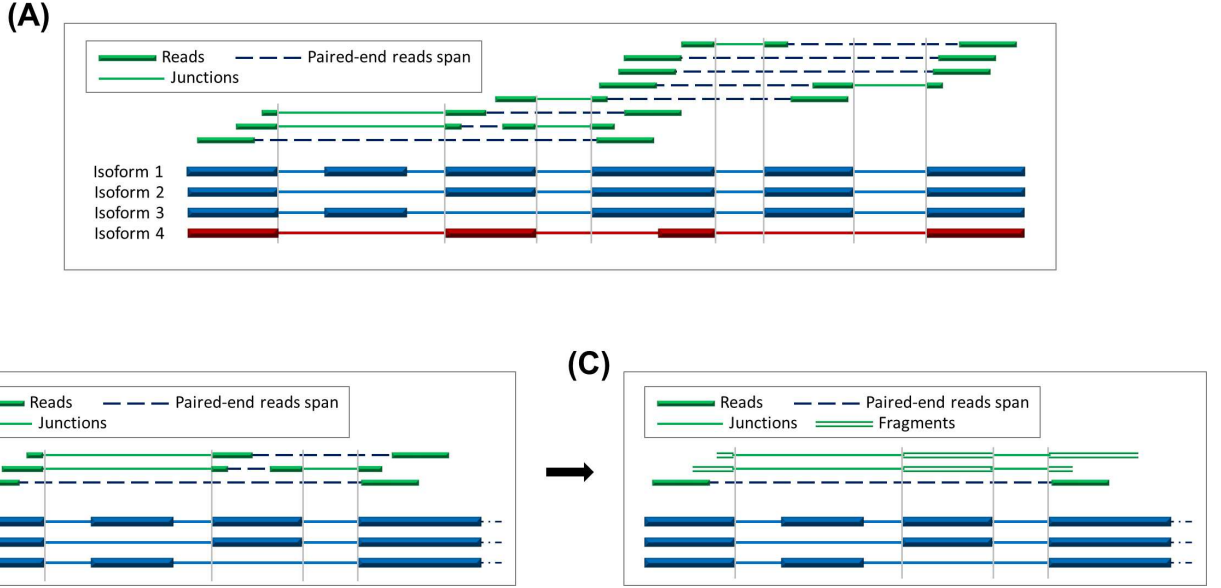}
\caption{An example of transcript filtering and signal augmentation. The upper panel (A) shows an example of how read counts information can help to filter out inactive isoforms.  The isoforms are in blue and red whereas the reads are in green. The junctions of a read are connected by the green lines while the read pairs are connected by blue dash lines. The $4^{th}$ isoform is not supported by any read and thus will be excluded. The lower panels (B, C) show an example of how to recover the original fragments where paired-end mates are separated by a gap. (B) The $1^{st}$ and $2^{nd}$ paired-end reads (from the top) is compatible with the $2^{nd}$ isoform and must skip the $2^{nd}$ exon while spanning the $1^{st}$, $3^{rd}$ and $4^{th}$ exons. The $3^{rd}$ paired-end read may originate from any of the three transcripts. (C) The exonic coverage signal is augmented for the $1^{st}$ and $2^{nd}$ paired-end reads, but not for the $3^{rd}$ paired-end reads since we can not determine whether the cDNA fragment covers the $2^{nd}$ and $3^{rd}$ exons.}
\label{fig:pairedend}
\end{figure*}

As the first pre-processing step, we count the number of compatible reads (single-end or paired-end) for each isoform.
Any isoform that has an potential expression abundance lower than a cutoff is filtered out. The potential expression of an isoform is calculated as the number of its compatible reads multiplied by the read length divided by the isoform length. We set the cutoff as low as $1\%$ of the gene's average coverage to avoid filtering out active isoforms. For example, in Figure~\ref{fig:pairedend}A, the $4^{th}$ isoform will be filtered out because it is not supported by any paired-end read.

As the second pre-processing step, we augment the per-base exonic coverage signal by assuming a paired-end read originates from a cDNA fragment. Our goal is to reconstruct the original cDNA fragment sequences and use them to estimate the expression abundance of each transcript. It is because a transcript abundance is more proportional to the total bases of its cDNA fragment than to the total bases of its sequenced reads. As such, when a pair of short reads overlap, the source cDNA fragment sequence is reconstructed by removing the overlapping signal. Otherwise, recovering the cDNA fragment sequence may be more complicated depending on the structure of the compatible isoforms.

Figure~\ref{fig:pairedend}B and \ref{fig:pairedend}C present a showcase example in which two cDNA fragment sequences can be unambiguously determined. The $1^{st}$ and $2^{nd}$ paired-end reads (from the top) must skip the $2^{nd}$ exon and span across the other three exons because they belong to the $2^{nd}$ transcript. In this case, we extend the short reads along the exonic regions to recover their cDNA fragment sequence. The $3^{rd}$ paired-end read may be from any of the three transcripts. Thus, the original cDNA fragment may or may not cover the exons in the middle. In this case, we just use the signal observed from the reads and do not augment signal. After this step, the remaining isoforms and the augmented per-base signal will serve as input for our shrinkage approach, which will be described in the following sections.

\subsection{An Informed Shrinkage Motivated by Naked Exons}
We have recently discovered a ubiquitous phenomenon that a large portion of exonic regions have no coverage signal. We denote these uncovered exonic regions as naked exons. We illustrate this phenomenon in Figure~\ref{fig:UncoveredRegions} using a real-world RNA-Seq data set, where the transcriptome of a liver tissue sample from a $37$ year-old Caucasian male was sequenced (NCBI SRA accession number ERX011211). We binarize the genes according to the number of mapped reads. In each bin, we plotted both the total number of genes (blue) and genes with at least $10\%$ naked exons (green). Clearly, a vast majority of genes ($88\%$), including highly expressed genes, possess naked exons. The similar trend was also observed for many other RNA-Seq data sets, generated using different Illumina platforms, from NCBI SRA. Motivated by the ubiquitous naked exons, we developed an informed shrinkage approach that shrinks the proportions of those isoforms not supported by the naked exons, and preserve active isoforms from the genes with low expression. The rationale is that isoforms with naked exons in a highly expressed gene are more likely to be inactive than those without.

\subsubsection{Gene Structure Model with Informed Shrinkage}
For each gene, we denote the number of bases in the exonic regions by $M$, the number of transcripts by $N$ and the gene-level expression abundance by $r$. Thus, the vector of observed per-base signal is given by $\mathbf{e} = [e_{1}, e_{2},..., e_{M}]^T$, where $e_{i}$ is the observed read coverage at the $i^{th}$ exonic position. We also denote the vector of transcript proportions by $\mathbf{p} = [p_{1}, p_{2},..., p_{N}]^T$, where $p_{j}$ is the abundance proportion of the $j^{th}$ transcript, $\sum_{j=1}^N p_{j} = 1$ and $0 \leq p_{j} \leq 1$  for all $j$. The transcript isoform structures are represented using a splicing matrix, $\mathbf{S} \in \{0,1\}^{M \times N}$ of $0$'s and $1$'s. Each row $i$ represents a single base, and each column $j$ represents a transcript isoform. $S_{ij}$=1 if the $j^{th}$ transcript isoform contributes to the exonic signal observed at the $i^{th}$ base and $S_{ij}$=0 otherwise.

\begin{figure*}[t]
\begin{center}
\includegraphics[width=0.48\textwidth]{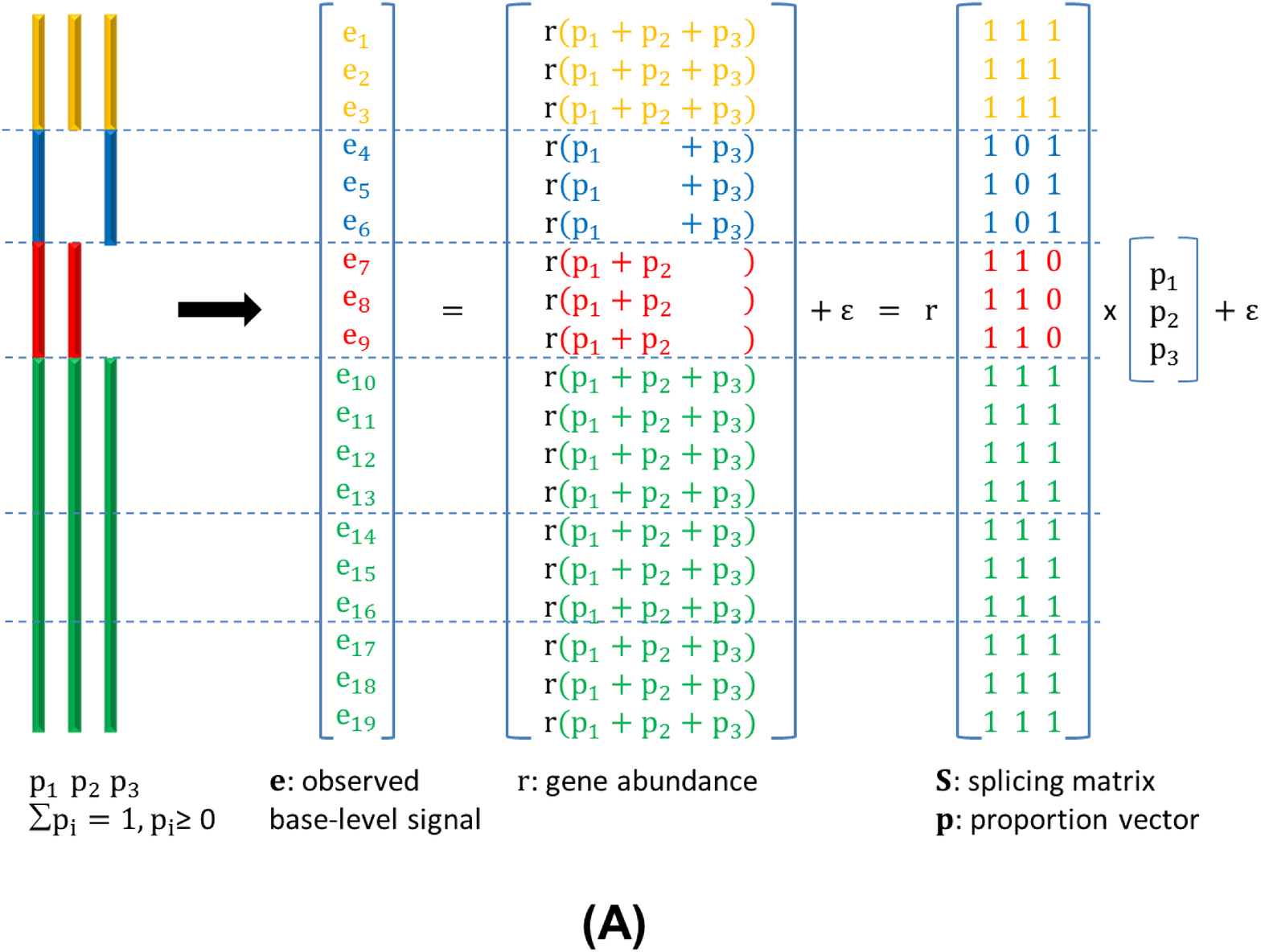}
\includegraphics[width=0.48\textwidth]{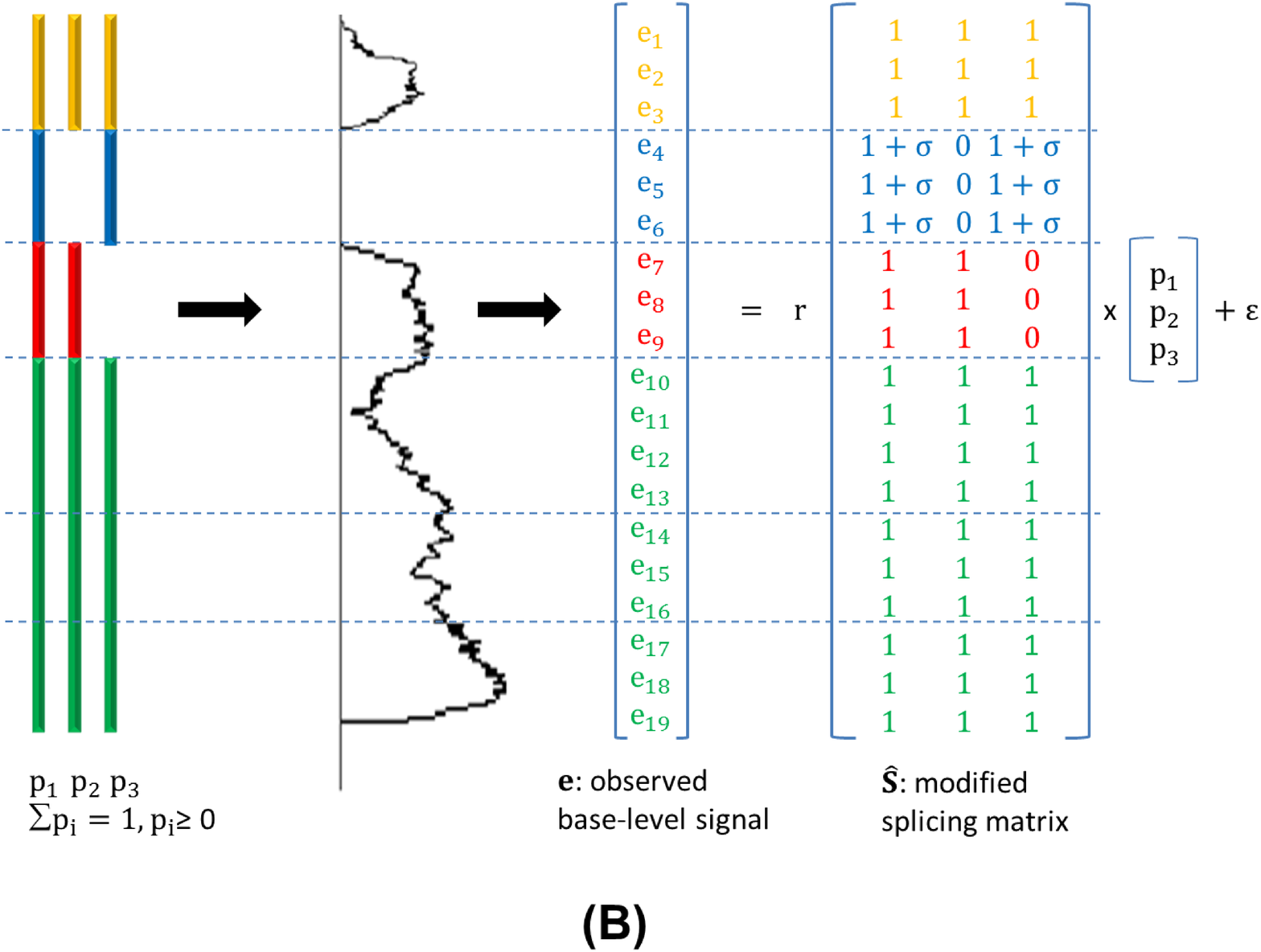}
\end{center}
\caption{Transcript quantification using the observed per-base exonic expression signal. $r$ represents the gene expression abundance parameter, $\mathbf{p}$ represents the abundance proportions of the three transcript isoforms, $\mathbf{e}$ represents the observed per-base exonic expression signal, $\mathbf{S}$ represents the splicing matrix, and $\mathbf{\hat{S}}$ represents the modified splicing matrix. In this example, the gene has three reference transcripts and four annotated exons. The $2^{nd}$ transcript skips the $2^{nd}$ exon whereas the $3^{rd}$ transcript skips the $3^{rd}$ exon. (A) The model without shrinkage. Each scalar $e_i$ corresponds to the expression signal of the $i^{th}$ exonic position and is expected to be the sum of transcript expression covering position $i$. (B) The selective and adaptive shrinkage model. A tuning parameter $\sigma$ is introduced to penalize selected regions (second exon) of the selected transcripts (the $1^{st}$ and the third) without an exonic expression signal. The shrinkage level is adaptively adjusted and optimized according to the exonic expression level over the algorithm iterations.}
\label{LeastSquare}
\end{figure*}

Figure~\ref{LeastSquare}A illustrates an example gene model without shrinkage. In this example, the $1^{st}$ transcript is full-length. Thus, all of the elements of the $1^{st}$ column of $\mathbf{S}$ take the value of $1$ ($S_{j1} = 1$ for all $j \in [1 \ldots M]$). In the $2^{nd}$ column of $\mathbf{S}$, the elements corresponding to the $2^{nd}$ exon take the value of $0$ because the $2^{nd}$ transcript skips the $2^{nd}$ exon. In the $3^{rd}$ column of $\mathbf{S}$, the elements corresponding to the $3^{rd}$ exon take the value of $0$ because the $3^{rd}$ transcript skips the $3^{rd}$ exon.

At each exonic position $i$, the expected per-base exonic expression signal is $r\sum_{j=1}^N S_{ij}p_{j}$ whereas the observed exonic expression signal is $e_i$. The latter is accumulated from both exonic and junction read coverage. In addition to exonic reads, junction reads are indispensable for identifying and quantifying adjacent exons as they cover the exon-exon junction regions. They are also essential for identifying short exons whose expression signal may be only identifiable by junction reads. Moreover, as the read length increases, a read is more likely to span multiple exons and hence carries valuable coverage information for our calculation. We write the observed per-base expression signal as a sum of the model explained portion and error as follows:
\begin{eqnarray*}
\left[\begin{array}{c}
e_1\\
e_2\\
%e_3\\
...\\
e_M
\end{array}\right]
&=&
\left[\begin{array}{c}
r(S_{11}p_1 + S_{12}p_2 + \ldots + S_{1N}p_N)\\
r(S_{21}p_1 + S_{22}p_2 + \ldots + S_{2N}p_N)\\
%r(S_{31}p_1 + S_{32}p_2 + \ldots + S_{3N}p_N)\\
\ldots\\
r(S_{M1}p_1 + S_{M2}p_2 + \ldots + S_{MN}p_N)
\end{array}\right] + \varepsilon\\
&=&
r\left[\begin{array}{cccc}
S_{11} & S_{12} & \ldots & S_{1N}\\
S_{21} & S_{22} & \ldots & S_{2N}\\
%S_{31} & S_{32} & \ldots & S_{3N}\\
\ldots\\
S_{M1} & S_{M2} & \ldots & S_{MN}
\end{array}\right] \times
\left[\begin{array}{c}
p_1\\
p_2\\
...\\
p_N
\end{array}\right] + \varepsilon,\\
\end{eqnarray*}
which we can rewrite in vector form as
\begin{equation}
\mathbf{e}= r\mathbf{S} \mathbf{p} + \varepsilon.
\label{eq:exp}
\end{equation}

For each gene locus, the relationship between the observed per-base exonic signal and the latent transcript proportions can be mathematically modeled as in Equation (\ref{eq:exp}), where $\varepsilon$ is the error term. In other words, we solve the following constrained linear least square problem:

\begin{equation}
\left\{
\begin{aligned}
 \left.\min_\mathbf{p} \|\mathbf{e}-r\mathbf{S}\mathbf{p}\|^2 \right.\\
 \mbox{subject to } \left.\mathbf{1}^{T} \mathbf{p} = 1, \right.\\
 \left.0 \leq \mathbf{p} \leq 1, \right.
\end{aligned}
\label{eq:single}
\right.
\end{equation}
where $\mathbf{1}^{T} = \underbrace{[1,1,1,...1]}_{\mbox{N}}$.

Our goal is to minimize $\varepsilon$ while avoiding overfitting. Biologically, since only a subset of transcripts are expressed under a single condition, it is important to avoid a possible model overfitting that uses the entire set of transcripts to explain the observed the exonic signal. For example, in Figure~\ref{LeastSquare}B, it is very likely that the reads originated from the $2^{nd}$ transcript. Otherwise, there should be exonic signal from the $2^{nd}$ exon (blue) shared by the $1^{st}$ and $3^{rd}$ transcripts. Simply solving the constrained linear least squares problem above will include the $3^{rd}$ transcript to explain the tower in the last region (green) as seen in Figure~\ref{LeastSquare}B.

We develop a selective and adaptive shrinkage approach through modifying the splicing matrix according to the observed exonic signal. Denoting $\sigma$ as a tuning parameter, we propose a new selective and adaptive shrinkage model with the modified structure matrix $\mathbf{\hat{S}}$ as the following:

\begin{equation}
    \hat{S}_{ij} = \left\{
        \begin{aligned}
         &\left. 1 + \sigma  \> \quad \mbox{ if } S_{ij}=1 \mbox{ and }  e_i=0, \right.\\
         &\left. S_{ij} \quad \mbox{ otherwise}, \right.
        \end{aligned}
        \right.
    \label{eq:modified}
\end{equation}
for all $i \in [1 \ldots M]$ and $j \in [1 \ldots N]$, where $\sigma$ is set to be proportional with $r$. The default value of $\sigma$ is $r$. It is very likely that any transcript with large uncovered regions in a highly expressed gene is inactive. Thus, highly expressed genes are penalized more for each uncovered base. On the other hand, this approach allows us to minimally penalize genes with low expression abundance or small uncovered regions because the modified splicing matrix is almost equal to the original splicing matrix $\mathbf{S}$. It should be noted that the above formulation may be sensitive to noise because it penalizes the exonic regions with exactly zero coverage. As such, to accommodate for noise, we penalize regions having very low average coverage compared to the overall coverage of the whole gene. Please refer to the Appendix for more details about the noise handling.

We implement two algorithms to solve the above optimization problem: One-Step SASeq and Iterative SASeq. One-Step SASeq calculates the gene expression abundance $r$ by averaging the gene's exonic signal. It then estimates the transcript proportions by minimizing the error term. On the other hand, Iterative SASeq iteratively estimates both the gene expression abundance and the transcript proportion vector simultaneously. While more accurate than One-Step SASeq, Iterative SASeq is more computationally intensive.

\subsubsection{One-Step SASeq} \label{sec:oneStep}
In One-Step SASeq, we first calculate $r$ by averaging the gene's exonic signal. We then rewrite the objective function in (\ref{eq:single}) as $(\mathbf{e} - r\mathbf{\hat{S}} \mathbf{p})^T (\mathbf{e} - r\mathbf{\hat{S}} \mathbf{p}) = \mathbf{p}^T (r^2 \mathbf{\hat{S}}^T \mathbf{\hat{S}}) \mathbf{p} - 2\mathbf{p}^T(r \mathbf{\hat{S}}^T\mathbf{e}) + \mathbf{e}^T \mathbf{e}$. After dropping the constant term $\mathbf{e}^T \mathbf{e}$, canceling out $r$ from the objective function and denoting $\mathbf{A} = r \mathbf{\hat{S}}^T \mathbf{\hat{S}}$ and $\mathbf{b} = \mathbf{\hat{S}}^T\mathbf{e}$, (\ref{eq:single}) takes the following form:

\begin{equation}
\left\{
\begin{aligned}
 %\left.\min_{\mathbf{p}} f(\mathbf{p}) = \frac{1}{2}\mathbf{p}^T (r \mathbf{\hat{S}}^T \mathbf{\hat{S}}) \mathbf{p} - \mathbf{p}^T( \mathbf{\hat{S}}^T\mathbf{e}), \right. \\
 \left.\min_{\mathbf{p}} f(\mathbf{p}) = \frac{1}{2}\mathbf{p}^T \mathbf{A} \mathbf{p} - \mathbf{p}^T\mathbf{b}, \right. \\
 \mbox{subject to } \left.\mathbf{1}^{T} \mathbf{p} = 1, \right.\\
 \left.0 \leq \mathbf{p} \leq 1, \right.
\end{aligned}
\label{eq:QP}
\right.
\end{equation}
where $\mathbf{A}$ is symmetrical and positive semidefinite. Please note that $\mathbf{A}$ and $\mathbf{b}$ are known and only $\mathbf{p}$ is the subject to be optimized. The above optimization problem can be solve by a classic convex quadratic programming. Please refer to the Appendix for more details about the active-set method for convex quadratic programming~\cite{Nocedal}.

\begin{figure*}[t]
\centering
\includegraphics[width=0.48\textwidth]{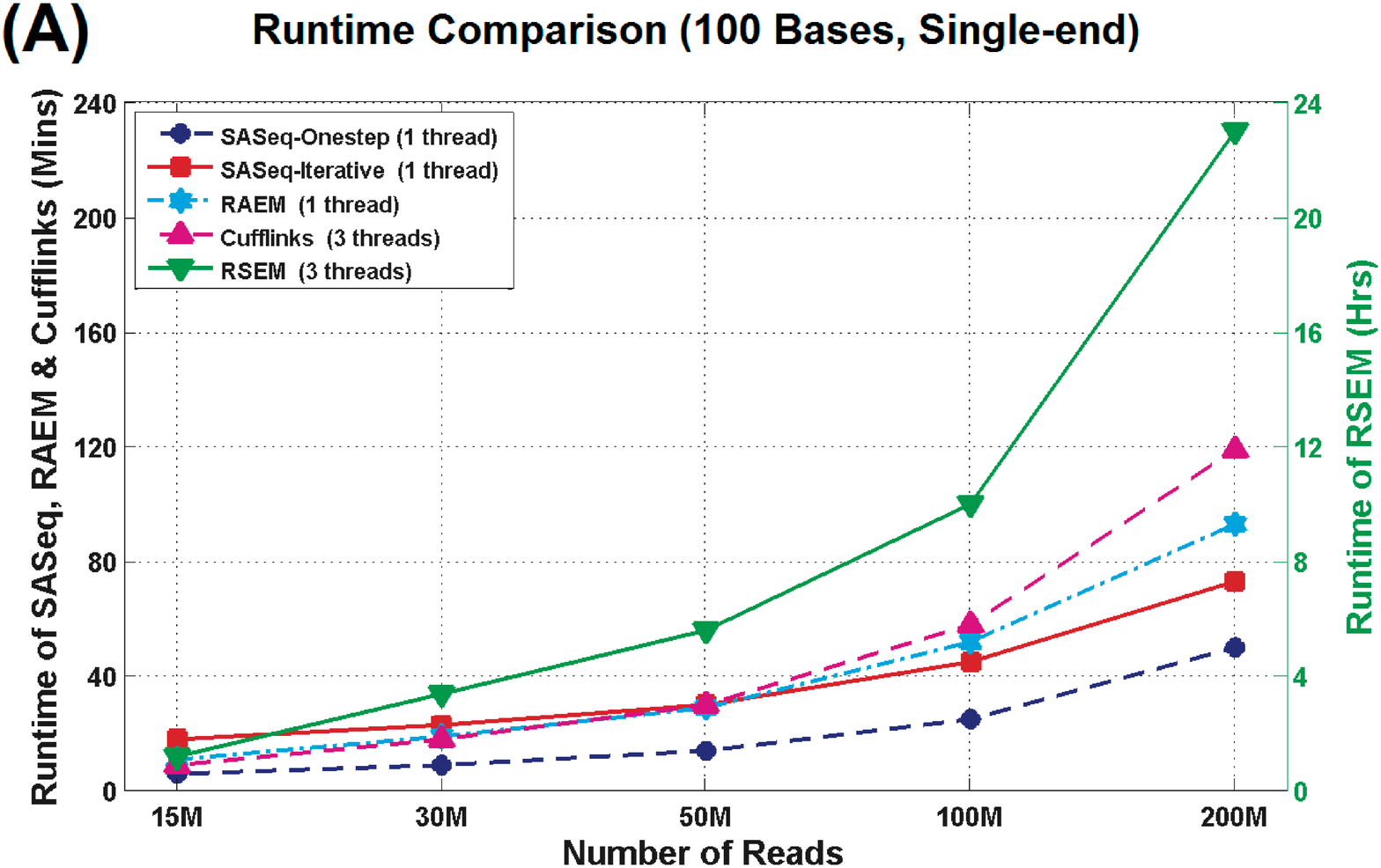}
\includegraphics[width=0.48\textwidth]{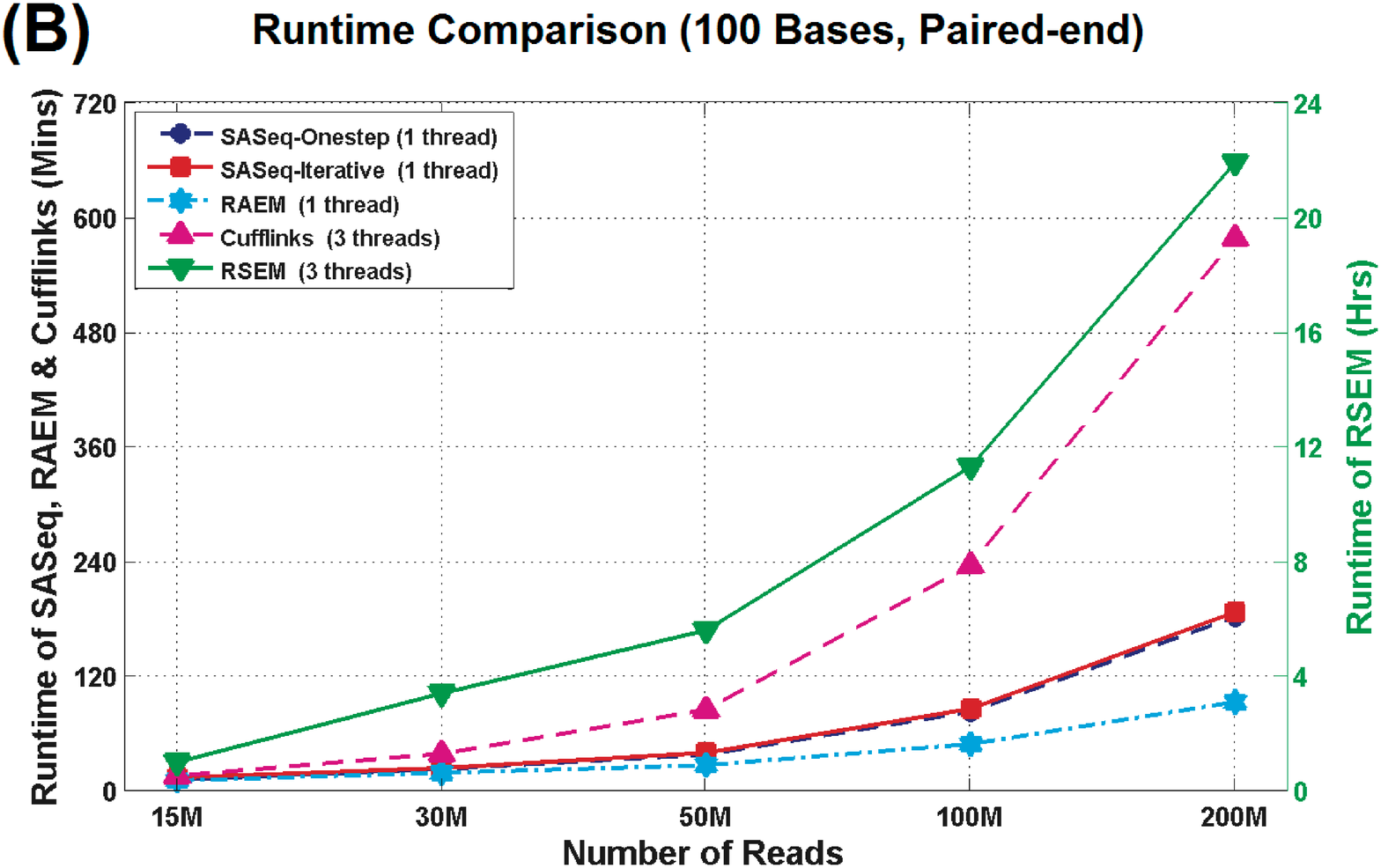}
\caption{Runtime comparisons of various algorithms. (A) The runtime comparison for single-end data sets. (B) The runtime comparison for paired-end data sets. For each plot, the horizontal axis represents the number of short reads in increasing order. The vertical axes represent the algorithm runtimes. For both data sets, RSEM is at the magnitude of hours (the right vertical axis) while all other algorithms are at the magnitude of minutes (the left vertical axis).}
\label{fig:time}
\end{figure*}

\subsubsection{Iterative SASeq} \label{sec:iterative}
In Iterative SASeq, we estimate both the gene-level expression abundance and transcript proportions at each iteration. Denoting $r^{(k)}$ and $\mathbf{p}^{(k)}$ as the expression abundance and transcript proportion at the $k^{th}$ iteration, the expression abundance $r^{(0)}$ can be initialized by averaging the per-base signal. Each iteration consists of two steps: the first step recalculates the transcript proportions while the second step updates the expression abundance.

In the first step of the $(k+1)^{th}$ iteration, the splicing matrices are modified using the gene-level expression abundance $r^{(k)}$ as follows:
\begin{equation}
    \hat{S}_{ij}^{(k+1)} = \left\{
        \begin{aligned}
         &\left. 1 + \sigma^{(k)}  \> \quad \mbox{ if } S_{ij}=1 \mbox{ and } e_i = 0, \right.\\
         &\left. S_{ij} \quad \mbox{ otherwise}, \right.
        \end{aligned}
        \right.
    \label{eq:modifiedriter}
\end{equation}
for all $i \in [1 \ldots M]$, and $j \in [1 \ldots N]$, where $\sigma^{(k)}$ is adjusted according to the expression abundance $r^{(k)}$ as described in (\ref{eq:modified})
. Denoting $\mathbf{A}^{(k+1)} = r^{(k)}(\mathbf{\hat{S}}^{(k+1)})^T\mathbf{\hat{S}}^{(k+1)}$, $\mathbf{b}^{(k)} = (\mathbf{\hat{S}}^{(k+1)})^T\mathbf{e}$, we recalculate the proportion vector $\mathbf{p}^{(k+1)}$ by solving the following convex QP:

\begin{equation}
\left\{
\begin{aligned}
   \left.\mathbf{p}^{(k+1)} = \arg\min_{\mathbf{p}} \{\frac{1}{2}\mathbf{p}^T \mathbf{A}^{(k+1)} \mathbf{p} - \mathbf{p}^T\mathbf{b}^{(k)}\}, \right.\\
   \mbox{subject to } \left.\mathbf{1}^{T} \mathbf{p} = 1, \right.\\
   \left.0 \leq \mathbf{p} \leq 1. \right.
\end{aligned}
\label{eq:step1}
\right.
\end{equation}
In the second step, given the new proportion vector $\mathbf{p}^{(k+1)}$ from (\ref{eq:step1}), we have $\mathbf{e}=r\mathbf{\hat{S}}^{(k+1)} \mathbf{p}^{(k+1)} + \varepsilon$. To calculate $r^{(k+1)}$, we optimize $r$ as follows:

\begin{equation}
r^{(k+1)} = \arg\min_{r} \|\mathbf{e} - r\mathbf{\hat{S}}^{(k+1)} \mathbf{p}^{(k+1)}\|^2. %\notag
\label{eq:step2}
\end{equation}

By solving (\ref{eq:step2}) for scalar values of $r$, we obtain the following result:

\begin{equation}
r^{(k+1)} = \frac{{\mathbf{e}}^T \mathbf{\hat{S}}^{(k+1)} \mathbf{p}^{(k+1)}}{\|\mathbf{\hat{S}}^{(k+1)} \mathbf{p}^{(k+1)}\|^2}.
\end{equation}

The algorithm iterates between the steps described in (\ref{eq:modifiedriter}), (\ref{eq:step1}) and (\ref{eq:step2}) until the proportion vector and expression abundance values converge.

\section{Results} \label{sec:results}
\subsection{Simulation Experiments}
Using simulation studies, we demonstrate the accuracy and speed of SASeq by comparing with Cufflinks \cite{Trapnell2010}, RAEM \cite{Deng2011} and RSEM \cite{RSEM}. We used FluxSimulator [\url{http://flux.sammeth.net/index.html}] to simulate the whole transcriptome sequencing experiments. FluxSimulator takes reference transcript sequences as the input, randomly generates copy numbers for each transcript, fragments them, selects the ones of right sizes to sequence \textit{in silico} and finally outputs the sequencing reads. Computational approaches for isoform identification and quantification can thus be compared in regards to the simulation ground truth of transcript isoforms and their copy numbers. FluxSimulator was used to generate $5$ data sets consisting of $15$ million, $30$ million, $50$ million, $100$ million and $200$ million single-end (SE) reads with lengths of $100$ from around $130,000$ reference transcripts available in the Ensembl database (version 65) with various copy numbers. FluxSimulator was also used to generate $5$ data sets with the same amount of short reads and length for paired-end (PE) reads. % (Figures S3, S4)
Both the simulation data used for method comparisons as well as SASeq method are freely available at \url{http://sammate.sourceforge.net/}.

Figure~\ref{fig:time} shows the time complexity of the different algorithms across an increasing number of short reads on the same server (4 x Twelve-Core AMD Opteron 2.6GHz, 256GB RAM). For the single-end data sets the runtime of single-thread SASeq is quite comparable with the single-thread RAEM \cite{Deng2011} and the multi-thread Cufflinks \cite{Trapnell2010}. It is also much faster than the multi-thread RSEM \cite{RSEM}. For paired-end data sets the runtimes of SASeq and RAEM are still comparable and are much faster than RSEM and Cufflinks. Due to the additional iterations of Iterative SASeq, it is slightly slower than One-Step SASeq.

We proceed to compare the accuracy of transcript isoform quantification using the true copy numbers of all of the transcript isoforms from the simulated data sets. For this purpose, we used the Jensen-Shannon (JS) divergence since it can capture both linear and nonlinear relationships. Thus, the most accurate transcript isoform quantification algorithm will yield a vector of isoform proportions that is least divergent from the vector of ground truth proportions. In other words, a lower value for the JS divergence indicates a better performance.
\begin{figure*}[t]
\centering
\includegraphics[width=0.48\textwidth]{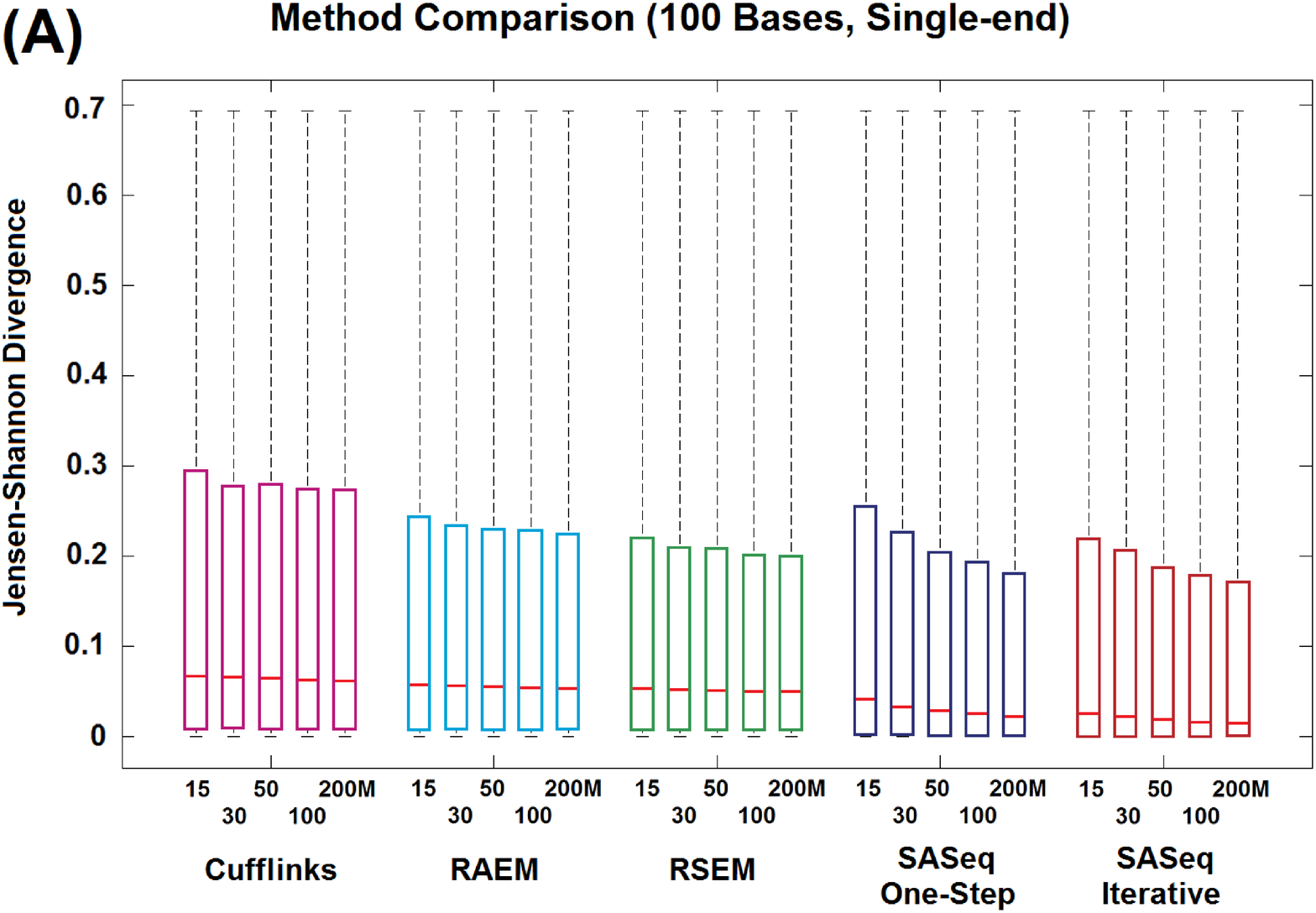}
\includegraphics[width=0.48\textwidth]{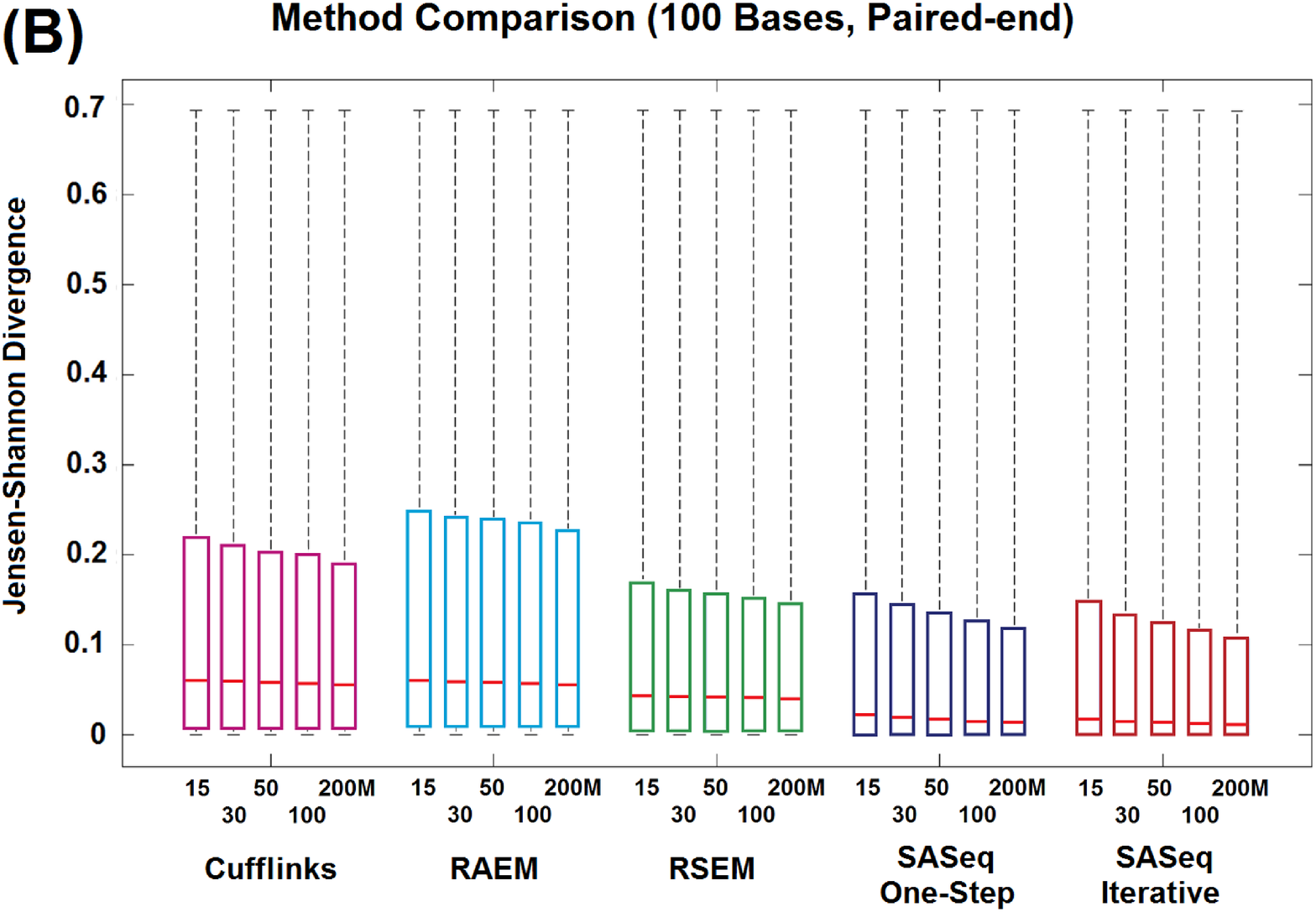}
\includegraphics[width=0.48\textwidth]{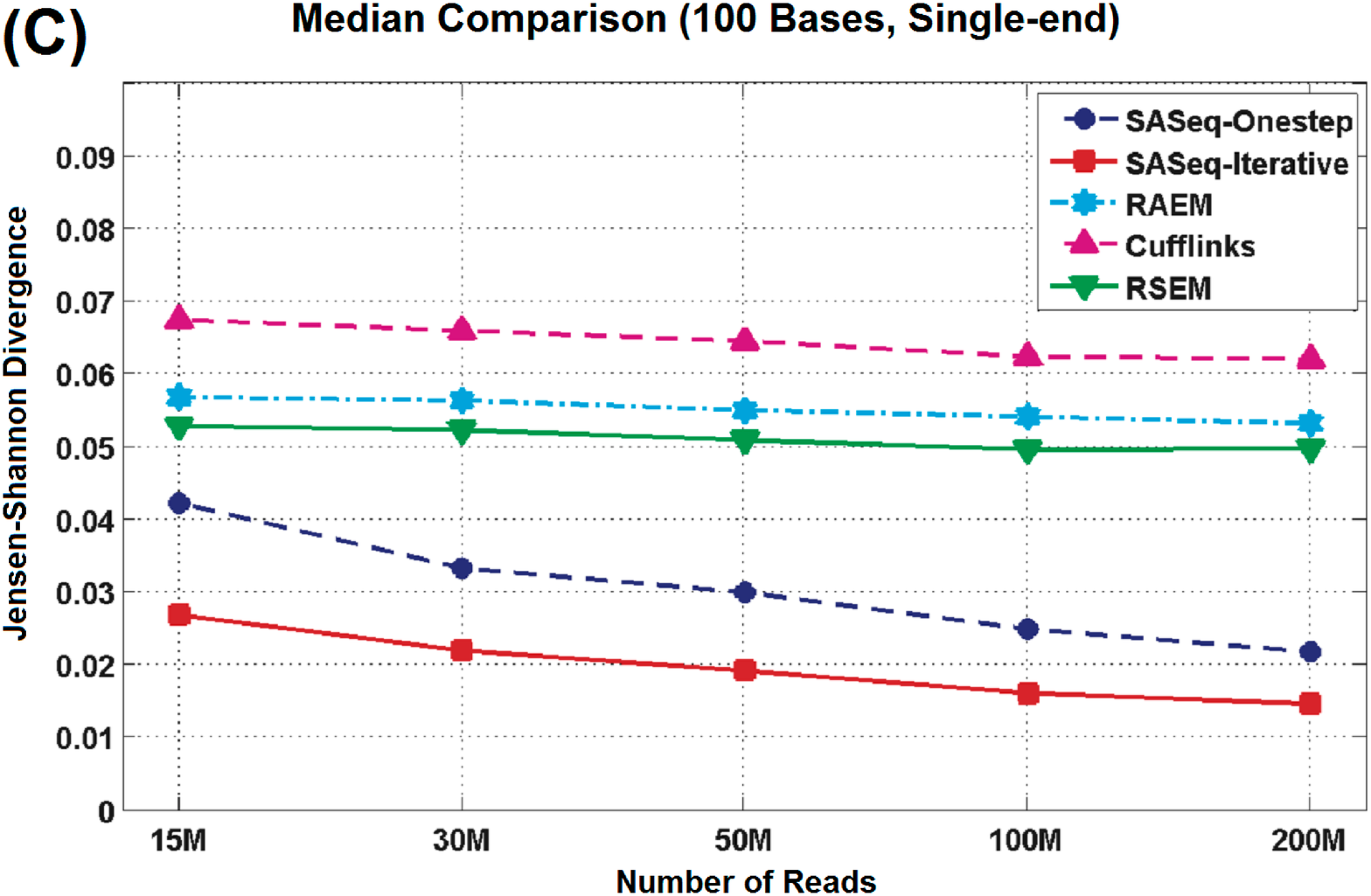}
\includegraphics[width=0.48\textwidth]{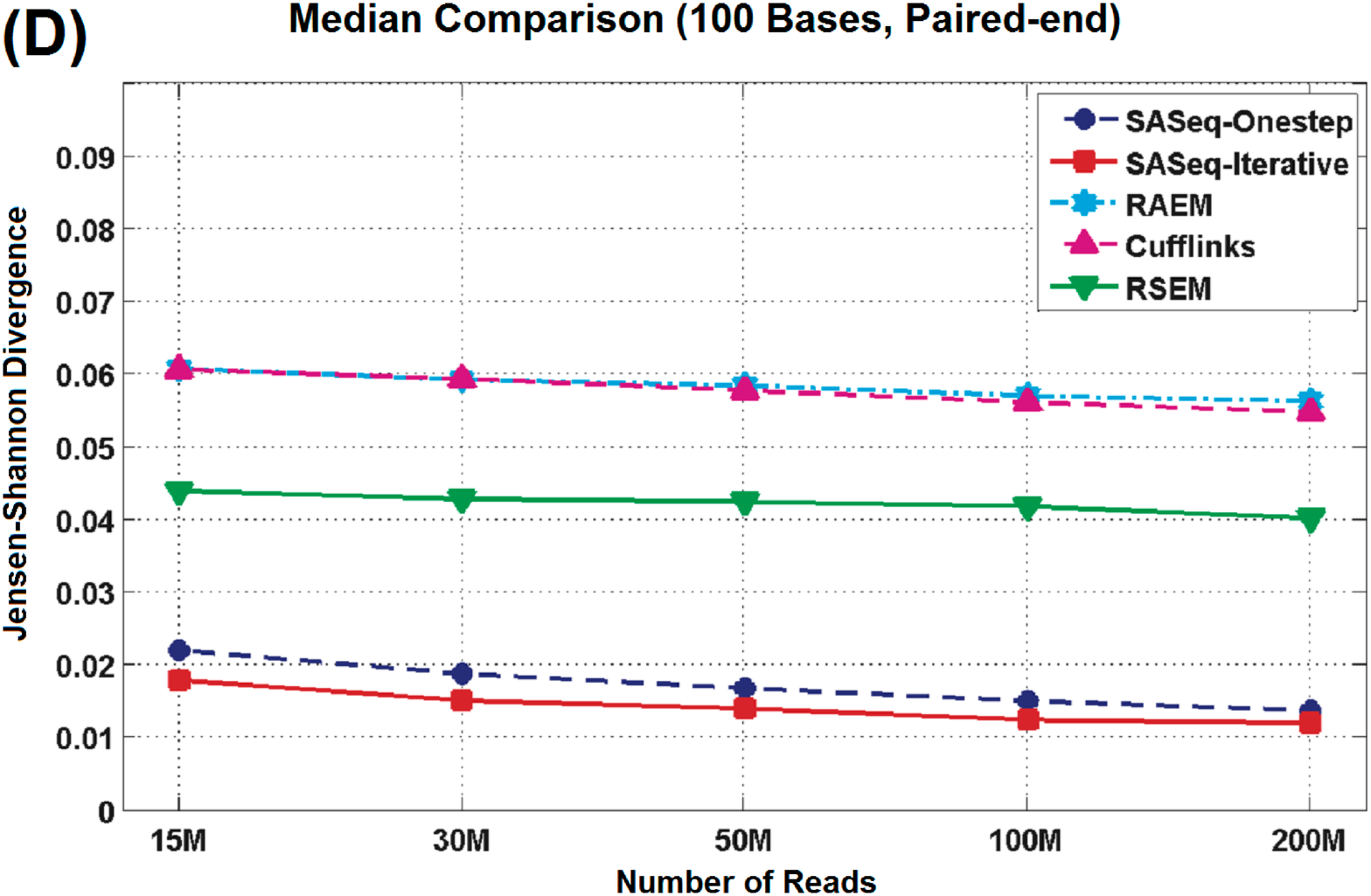}
\caption{Comparison of transcript quantification accuracy using JS divergence. The upper left panel (A) compares algorithms in terms of their divergence distributions from the ground truth for single-end data sets. The upper right panel (B) does the same for the paired-end data sets. The lower left panels (C, D) compare algorithms in terms of their medians from the ground truth for single-end and paired-end data sets, respectively. Both One-Step and Iterative SASeq algorithms outperform their competitors for both the single-end and paired-end data sets. The medians of SASeq are distinguishingly lower than of the competing methods. The performance contrasts get sharper as the number of short reads increases.
}\label{fig:JS}
\end{figure*}
In Figure~\ref{fig:JS}A and \ref{fig:JS}B, the horizontal axes represent an increasing number of short reads for the single-end and paired-end data sets, respectively. The vertical axes of Figure~\ref{fig:JS}A and \ref{fig:JS}B represent the JS divergence distributions for the aforementioned data sets. Lower values indicate a better performance. In Figure~\ref{fig:JS}C and \ref{fig:JS}D, we also plot the medians of each JS divergence distribution for the single-end and paired-end data sets, respectively. Please note that the number of paired-end reads equals half of the number of short reads shown in Figure~\ref{fig:JS}B and \ref{fig:JS}D. For example, the $200$M data set in Figure~\ref{fig:JS}B and \ref{fig:JS}D actually has $100$ million paired-end reads.

In general, all the methods perform better for the paired-end data sets except RAEM, which does not exploit paired-end information. For all data set, SASeq outperforms its competitors with Iterative SASeq being superior to One-Step SASeq. More strikingly, the performance gap between SASeq and other algorithms widens as the number of reads increases for both the single-end and paired-end data sets. Furthermore, the significantly smaller variances for the paired-end data sets indicate a robust behavior of SASeq. There are likely three reasons for the performance gaps between SASeq and other methods. First, other approaches do not exploit the ubiquitous phenomenon of uncovered regions. Second, the signal correction procedure helps to yield better per-base signal and thus enhances the numeric estimation of expression abundances of active transcripts. Third, combining available information from both read counts and per-base signal yield a much better result than using each of them alone. This is translated into a prospectively more powerful algorithm for the forthcoming ultra-high throughput sequencing data and longer read length. In summary, from Figure~\ref{fig:time} and Figure~\ref{fig:JS}, SASeq demonstrates an impressive accuracy at a reasonable speed.

\subsection{Real RNA-Seq Data}
\begin{figure*}[t]
\centering
\includegraphics[width=0.96\textwidth]{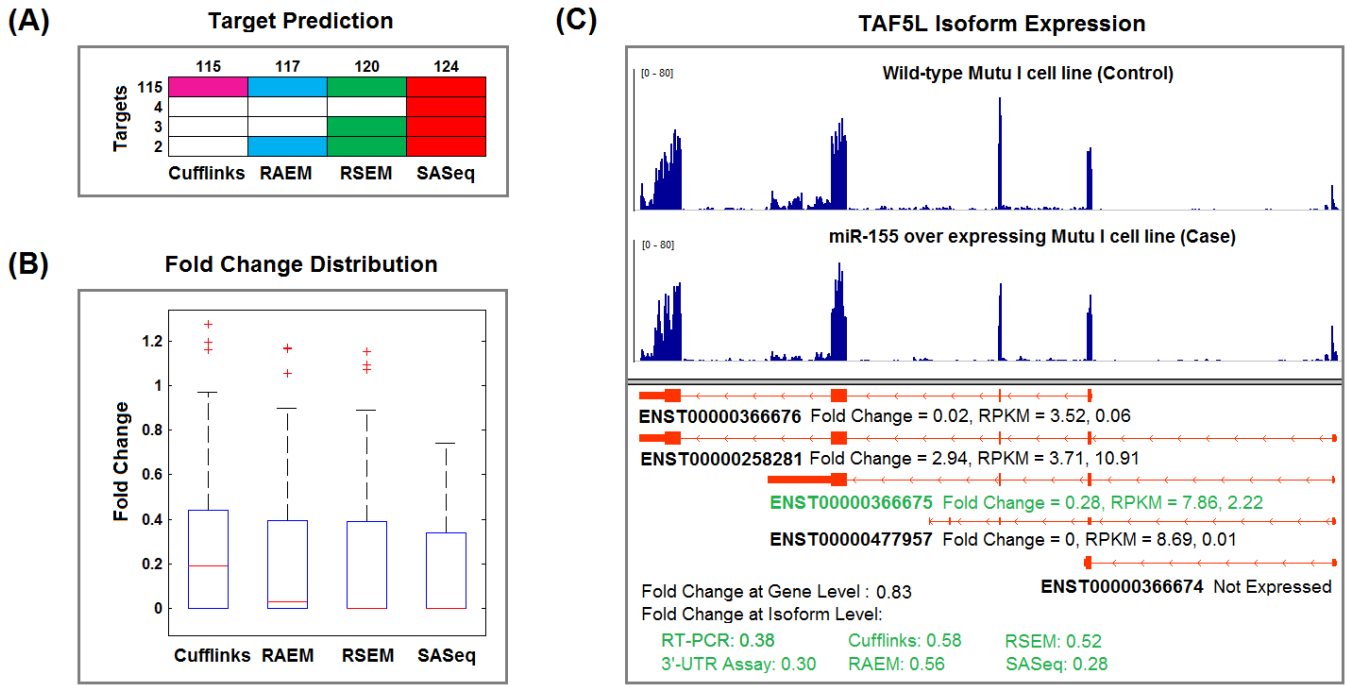}
\caption{Predicted mir-155 targets. (A) The comparison of target prediction using a bitmap. Numbers at the \emph{top} show the total predicted number of down-regulated genes of each method. Numbers to the \emph{left} show the number of genes common in each respective row. For example, the first row indicates that $115$ genes are predicted by all of the methods whereas the second row indicates that $4$ genes are uniquely predicted by SASeq. (B) The isoform-level fold change distributions of $124$ targets validated by 3'-UTR assay from the previous study \cite{Xu2010}. (C) Quantification of the gene TAF5L. SASeq predicts both ENST00000366675 and ENST00000366676 as down-regulated in the mir-155 expressing cell line with $0.83$ gene-level differential expression. The isoform \emph{ENST00000366675} has a mir-155 seed in the 3'-UTR, which was validated by both RT-PCR and 3'-UTR assay \cite{Deng2011}. The numeric values in green show the fold changes predicted by the competing methods and by both validations (RT-PCR and 3'-UTR assay) for the  \emph{ENST00000366675} isoform.}\label{target}
\end{figure*}

We also demonstrate the practical utilities of SASeq using a real RNA-Seq data set of six Mutu I wild-type cell line %samples
and six mir-155 expressing cell line samples (accession number: SRA011001). The wild-type Mutu I cell line is of type I latency (limited viral gene expression) B-cell line whereas the mir-155 expression cell line is introduced by infecting the Mutu I cell line in duplicate with an mir-155 expressing retrovirus (or an empty vector control retrovirus) to achieve a high mir-155 expression \cite{Xu2010}. The goal of the data analysis is to identify the down-regulated transcripts that can be potential mir-155 targets. We picked this RNA-Seq data set because hundreds of gene expression were quantified and a large number of 3'-UTR assays have been done to validate the predicted mir-155 targets \cite{Deng2011,Xu2010}. We first performed a per-base sequence quality check using fastQC Software (http://www.bioinformatics.bbsrc.ac.uk/projects/fastqc/).
We then used TopHat \cite{TopHat} to uniquely align short reads to the human reference genome (hg19/GRCh37). Default settings and unique alignment were used. Alignment results were saved in a BAM format. We re-analyzed the data set using SASeq to re-discover the previously reported differentially expressed transcripts and to predict novel targets.

We compared the performance of Cufflinks, RAEM, RSEM, and SASeq in accurately identifying down-regulated transcript isoforms. We used a ``gold standard" data set of mir-155 targets (genuine down-regulation) validated by 3'-UTR assays \cite{Xu2010}. We pooled the short reads from the six samples within each condition to gain more exonic coverage for the target genes. We also removed $25\%$ of the transcripts that were least expressed in the control case to obtain reliable fold changes. We then conducted isoform-level differential expression analysis to screen down-regulated transcript isoforms of the $124$ mir-155 targets that were validated by 3'-UTR assay in the previous study \cite{Xu2010}. Using the same cutoff value of $0.8$ as in the aforementioned studies \cite{Deng2011,Xu2010}, SASeq is able to predict more down-regulated genes than other competing methods (Figure~\ref{target}A). We also plotted the predicted fold changes of $124$ mir-155 targets in Figure~\ref{target}B, which shows that the fold changes predicted by SASeq are more significant than other methods.

Particularly, we demonstrate in Figure~\ref{target}C that SASeq identifies the down-regulated mir-155 target at the isoform-level, which has been validated by both RT-PCR and 3'-UTR assay. In Figure~\ref{target}C, the upper panel shows the exonic expression signal for the gene TAF5L whereas the lower panel shows the structures of five reference transcript isoforms and the associated data analysis. A fold change of $0.83$ as well as a visual inspection of the upper panel show no significant differential expression at the gene-level. Furthermore, the target transcript isoform ENST00000366675 has been validated by both RT-PCR and 3'-UTR assay with fold changes of $0.38$ and $0.30$, respectively \cite{Deng2011}. Using SASeq we predicted the most accurate fold change of $0.28$ compared to Cufflinks' ($0.58$), RAEM's ($0.56$), and RSEM's ($0.52$).

\section{Conclusions} \label{sec:conclusion}

In this paper, we presented a new approach to identify and quantify active transcripts using RNA-Seq. Our contribution is two-fold. First, we combine the two types of complementary information available, i.e. the read counts and the per-base signal, to yield a more accurate result. Second, we develop a novel shrinkage technique to shrink transcript proportions that are not supported by the observed per-base exonic expression signal and to adaptively adjust the shrinkage level accordingly. Since the ever-increasing numbers of reference transcripts in the databases as well as their error rates aggravate the risk of model overfitting, model selection via shrinkage is a promising avenue for future research in this area.

The key innovation of our approach is the informed shrinkage, which is fundamentally different from a Lasso shrinkage approach. First, choosing the value of the tuning parameter for the standard Lasso shrinkage approach is not straightforward because the number of condition-specific transcripts is unknown. Furthermore, this number varies from gene to gene and from condition to condition. On the other hand, SASeq automatically determines the value of its tuning parameter according to the overall exonic coverage signal of the gene. Finally, for any gene the number of condition-specific transcripts is usually very small compared to the number of reference transcripts available from a database. With that in mind, the Lasso approach shrinks all the reference transcript abundance parameters in a non-discriminative way and may not necessarily lead to a set of condition-specific transcripts. On the other hand, SASeq only penalizes selected regions of the transcripts that are not supported by naked exons, the remaining is more likely to be expressed under a specific biological condition.

Nevertheless, SASeq will prospectively become more powerful in the near future with the accelerated augmentation of transcript databases and increasing sequencing depths. The former will give a more complete set of transcripts to select from. The latter will permit a more accurate selection and shrinkage of the transcripts and their regions. In addition to working with transcript databases, SASeq is able to identify and quantify active transcripts from transcriptome assembly outputs in the gtf format \cite{Trapnell2010,Trinity,TransAbyss,DBLP:conf/bibm/ZhaoNDJZ11,Oases2012} where model misspecification is also an outstanding issue due to the excessive assembly bias and errors. SASeq is freely available via the GUI software SAMMate at \url{http://sammate.sourceforge.net/}.

% I would discuss about downstream analysis after quantification (differential analysis and network inference) but I'm afraid that I don't have enough knowledge to give a strong discussion

\section{Appendix} \label{sec:appendix}
\subsection{Noise Handling}\label{sec:AverageNoise}

\begin{figure}[t]
\centering
\includegraphics[width=0.48\textwidth]{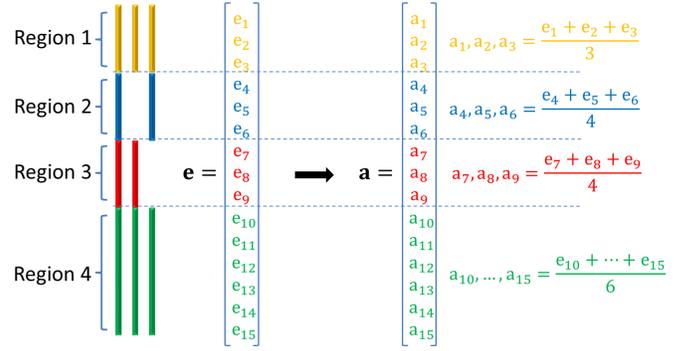}
\caption{The average coverage matrix.}
\label{fig:AverageNoise}
\end{figure}

To accommodate for noise, we consider the average coverage of each exonic regions. We first calculate the average coverage signal over each region and then penalize those regions having very low average coverage compared to the overall coverage of the whole gene (Figure~\ref{fig:AverageNoise}). Define $\mathbf{a} =  [a_{1}, a_{2},..., a_{M}]^T$ as a new vector, each scalar value $a_i$ is the average coverage of the region that the $i^{th}$ base belongs to. For One-Step SASeq, we change the Equation (\ref{eq:modified}) as follows:

\begin{equation}
    \hat{S}_{ij} = \left\{
        \begin{aligned}
         &\left. 1 + \sigma  \> \quad \mbox{ if } S_{ij}=1 \mbox{ and } a_i<\alpha, \right.\\
         &\left. S_{ij} \quad \mbox{ otherwise}, \right.
        \end{aligned}
        \right.
    \label{eq:NoiseModified1}
\end{equation}
for all $i \in [1 \ldots M]$ and $j \in [1 \ldots N]$, where $\alpha$ is the new signal-noise cutoff parameter with a default value of $r/100$.

Similarly, for Iterative SASeq, we change the Equation (\ref{eq:modifiedriter}) as follows:

\begin{equation}
    \hat{S}_{ij}^{(k+1)} = \left\{
        \begin{aligned}
         &\left. 1 + r^{(k)}  \> \quad \mbox{ if } S_{ij}=1 \mbox{ and } a_i<\alpha^{(k)}, \right.\\
         &\left. S_{ij} \quad \mbox{ otherwise}, \right.
        \end{aligned}
        \right.
    \label{eq:NoiseModified2}
\end{equation}
for all $i \in [1 \ldots M]$, and $j \in [1 \ldots N]$, where $\alpha^{(k)}$ is adjusted according to the expression abundance $r^{(k)}$ as described in (\ref{eq:NoiseModified1}).

\subsection{Convex Quadratic Programming Algorithm}\label{sec:algorithmQP}
We use the active-set method \cite{Nocedal} to solve the convex QP described above. Our convex QP can be written as:

\begin{equation}
\left\{
\begin{aligned}
 \left.\min_{\mathbf{p}} f(\mathbf{p}) = \frac{1}{2}\mathbf{p}^T \mathbf{A} \mathbf{p} - \mathbf{p}^T \mathbf{b}, \right. \\
 \mbox{subject to } \left.\mathbf{1}^{T} \mathbf{p} = 1, \right.\\
 \left.0 \leq \mathbf{p} \leq 1, \right.
\end{aligned}
\label{eq:CQP}
\right.
\end{equation}
where $\mathbf{A}$ is symmetric and positive semidefinite $N\times N$ matrix, $\mathbf{b}$ and $\mathbf{p}$ are vectors with $N$ elements. This convex QP has one equality constraint and $2N$ inequality constraints.

\noindent {\bf Algorithm} Active-Set Method for Convex QP
\begin{algorithmic}
\STATE Set $\mathbf{p}^{(0)} = [1, 0, \ldots,0]^T$ as the feasible starting point;
\STATE Set $\mathcal{W}^{(0)}$ to be a subset of the active constraints at $\mathbf{p}^{(0)}$;
\STATE Set $k=0$;
\LOOP
    \STATE Find $\Delta \mathbf{p}^{(k)}$ to minimize $f(p^{(k)} + \Delta \mathbf{p}^{(k)})$ in the subspace defined by $\mathcal{W}^{(k)}$;
    \IF {($\Delta \mathbf{p}^{(k)}==0$)}
        \STATE Compute Lagrange multipliers of inequality-constraints included in $\mathcal{W}^{(k)}$;
        \IF {(exists a negative Lagrange multiplier)}
            \STATE Obtain $\mathcal{W}^{(k+1)}$ by dropping the corresponding constraint from $\mathcal{W}^{(k)}$;
            \STATE Set $\mathbf{p}^{(k+1)} = \mathbf{p}^{(k)}$;
        \ELSE
            \RETURN $\mathbf{p}^{(k)}$;
        \ENDIF
    \ELSE
        \IF {($p^{(k)} + \Delta \mathbf{p}^{(k)}$ is feasible with respect to all constraints)}
            \STATE Set $\mathbf{p}^{(k+1)} = p^{(k)} + \Delta \mathbf{p}^{(k)}$;
            \STATE Set $\mathcal{W}^{(k+1)} = \mathcal{W}^{(k)}$
        \ELSE
            \STATE Set $\mathbf{p}^{(k+1)} = \mathbf{p}^{(k)} + \alpha^{(k)} \Delta\mathbf{p}^{(k)}$ where $\alpha^{(k)}$ is chosen to be the largest value for which all the constraints are satisfied;
            \STATE Obtain $\mathcal{W}^{(k+1)}$ by adding the blocking constraints to $\mathcal{W}^{(k)}$;
        \ENDIF
    \ENDIF
    \STATE Set $k=k+1$;
\ENDLOOP
\end{algorithmic}
\smallskip

For a given iteration $\mathbf{p}^{(k)}$ and a working set $\mathcal{W}^{(k)}$, if the objective function is not minimized in the subspace defined by the working set, we compute the step $\Delta\mathbf{p}^{(k)}$ by solving an equality-constrained QP, in which the constraints corresponding to the working set $\mathcal{W}^{(k)}$ are treated as equalities. We set $\mathbf{p}^{(k+1)} = \mathbf{p}^{(k)} + \Delta \mathbf{p}^{(k)}$ if it is feasible to all constraints. Otherwise, we set $\mathbf{p}^{(k+1)} = \mathbf{p}^{(k)} + \alpha^{(k)} \Delta \mathbf{p}^{(k)}$ where the step-length $\alpha^{(k)}$ is chosen to be the largest value for which all the constraints are satisfied. The blocking constraint is also added to the working set. If the objective function is minimized ($\Delta\mathbf{p}^{(k)}$=0) in the subspace defined by $\mathcal{W}^{(k)}$ but one of the Lagrange multipliers corresponding to the an inequality constraint in the working set is negative (does not satisfy the Karush-Kuhn-Tucker condition), we remove this constraint from the working set. Upon reaching a Karush-Kuhn-Tucker point that minimizes the objective function over its current working set, the algorithm terminates.

\section*{Acknowledgments}

The authors would like to thank Guorong Xu and Zhansheng Duan for their initial efforts on this work. The authors also thank Thair Judeh for his useful comments.

\bibliographystyle{nar_max_authors_10}
\bibliography{references}

\end{document}